# Observation of hybrid Tamm-plasmon exciton-polaritons with GaAs quantum wells and a MoSe$_2$ monolayer

Matthias Wurdack[1], Nils Lundt[1], Martin Klaas[1], Vasilij Baumann[1], A. Kavokin[2,3], Sven Höfling[1,4] and Christian Schneider[1]

[1]*Technische Physik and Wilhelm-Conrad-Röntgen-Research Center for Complex Material Systems, Universität Würzburg, D-97074 Würzburg, Am Hubland, Germany.*

[2]*Physics and Astronomy School, University of Southampton, Highfield, Southampton, SO171BJ,*
*UK*

[3]*SPIN-CNR, Viale del Politecnico 1, I-00133 Rome, Italy*

[4]*SUPA, School of Physics and Astronomy, University of St. Andrews, St. Andrews KY 16 9SS, United Kingdom*

**Strong light matter coupling between excitons and microcavity photons, as described in the framework of cavity quantum electrodynamics, leads to the hybridization of light and matter excitations. The regime of collective strong coupling arises, when various excitations from different host media are strongly coupled to the same optical resonance. This leads to a well-controllable admixture of various matter components in three hybrid polariton modes. Here, we study a cavity device with four embedded GaAs quantum wells hosting excitons that are spectrally matched to the A-valley exciton resonance of a MoSe$_2$ monolayer. The formation of hybrid polariton modes is evidenced in momentum resolved photoluminescence and reflectivity studies. We describe the energy and k-vector distribution of exciton-polaritons along the hybrid modes by a thermodynamic model, which yields a very good agreement with the experiment.**

**Introduction**

The regime of strong coupling between confined excitons and microcavity photons was first observed in 1992 by Weisbuch et al. in a semiconductor microcavity with three embedded quantum wells [1]. From the fundamental point of view, the light-matter hybridization allows to create quasi-particles with precisely tailored properties, such as well-defined effective masses and interaction constants [2]. The fundamental hybrid excitations in this device, the so-called exciton polaritons, have become a convenient laboratory for studies of collective quantum effects in semiconductors, including coherent phenomena such as Bose-Einstein condensation [3-5], topological excitations [6,7] and superfluidity [8]. Polariton condensates have been recently proposed as a platform for practical quantum emulation [9-11]. While a significant progress in the investigation of polaritonic effects in media with highly stable excitonic excitations such as organic materials [12-14] or GaN [15] has been reported, clearly, by far the most striking effects are observed in GaAs based structures at cryogenic temperatures so far. One additional particular property, which makes cavity polaritons an appealing system for quantum technologies is their spinor degree of freedom, leading to a variety of unique spin- and magnetic effects observed in linear and non-linear regimes [16-18]. While spin effects in GaAs quantum wells are rather fragile due to fast spin relaxation, the exciton pseudospin is much better preserved in monolayers of transition metal dichalcogenides (TMDCs) due to pseudospin-valley locking [19]. The formation of cavity polaritons based on excitons in TMDC monolayers [20-23] and van der Waals heterostructures [21] has been reported just recently. Subsequently, a device which features a hybrid polariton mode with an admixture of organic and monolayer excitons has been reported [24]. However, in the reported scenario, both host materials feature very strong exciton binding energies and small Bohr-radii, which prevents one from taking the full advantage of the organic-inorganic hybridization mechanism. Furthermore, electro-optical effects and current injection are difficult to implement in this kind of device.

Here, we study a hybrid structure, which hosts both large-radius excitons in GaAs quantum wells and tightly bound excitons in a MoSe$_2$ monolayer. Such a device represents a first crucial step towards combining the unique physics inherent to two-dimensional materials with the well matured device platform in III-V optoelectronics and polaritonics. We demonstrate that both types of excitations enter the strong coupling regime with the same cavity resonance, which gives rise to hybrid excitations that admix excitons of TMDC monolayers and conventional GaAs quantum wells. This new kind of quasiparticle is evidenced in temperature dependent angle-resolved photoluminescence (PL) and reflectivity experiments. Our experimental findings are supported by a theoretical formalism based on the two-coupled oscillator model, and the PL angular dependence is shown to follow the simple thermal distribution of polariton states.

**Results**

**Device Fabrication**

Fig. 1a depicts a graphic illustration of our microcavity device: It consists of an AlAs/AlGaAs distributed Bragg reflector (DBR) (30 pairs), which is characterized spectrally by its stopband ranging from 710 nm to 790 nm (Fig. 1b), supporting a reflectivity up to 99.9 % between 740 nm (1.675 eV) and 765 nm (1.621 eV) at 10 K. The AlAs/AlGaAs Bragg stack, which has been grown by Gas Source MBE (molecular beam epitaxy), is topped with a 112 nm thick AlAs layer with four embedded GaAs Quantum Wells (QWs) (details can be found in the Method section of the manuscript). A layer of GaInP caps the AlAs layer, to prevent surface oxidation.

Fig. 1b) shows the reflectivity and PL spectra of this device: We observe a strong PL peak, which we attribute to the emission from GaAs excitons at 749 nm. The resonance correlates with the characteristic absorption dip in the reflectivity spectrum. In addition, the PL spectrum features a second peak at approximately 760 nm, which only shows a marginal absorption dip in the reflectivity. Most likely, this feature corresponds to defect assisted transitions in the QWs with small oscillator strength.

As the next step, a single layer of $MoSe_2$, mechanically exfoliated via commercial adhesive tape from a bulk crystal was transferred onto the top GaInP layer with a polymer stamp. The monolayer was identified by its reflection contrast in our optical microscope prior to the transfer onto the heterostructure. Fig. 1c) shows the reflectivity spectrum recorded at the sample spot containing a TMDC monolayer. This reflectivity spectrum convolutes the absorption resonance of the A-valley exciton of $MoSe_2$ with the GaAs QW exciton resonance. By norming it with the reference spectrum recorded on a sample spot next to the monolayer (containing the bare QWs), we can recapture the characteristic MoSe$_2$ reflection spectrum, as shown in Fig. 1d). We note, that the spectrum is strongly dominated by the resonance corresponding to the neutral exciton transition, whereas the trion absorption is only weakly visible at 1.634 eV.

The full cavity device is completed by capping the monolayer with an 80 nm thick layer of polymethyl methacrylate (PMMA) and a 60 nm thick gold layer (Fig. 2a). These layers were carefully designed to support an optical resonance with a field antinode both at the position of the monolayer and at the stack of GaAs QWs, while ensuring the spectral resonance with both modes. A transfer matrix calculation of the reflectivity spectrum of our device with nominal layer thicknesses (see Methods section for details) is shown in Fig. 2b. The corresponding refractive indices of the layer sequence and the resulting optical mode profile are shown in Fig 2c. The resonance features a theoretically expected Q-factor of 1095 which is limited by the finite reflection of the metal top layer.

**Optical characterization**

We will first discuss the formation of exciton polaritons based on the GaAs QW exciton resonance, recorded on a sample spot a few micrometers next to the TMDC monolayer location. We employ the characteristic energy-momentum dispersion relation of the vertically confined photon field to map out the coupling of GaAs exciton and the cavity mode, which is recorded in the PL and reflectivity experiments in the far-field imaging configuration (see Method section for further details). The sample is studied at various temperatures between 10 K and 200 K, and the device is excited by a non-resonant continuous wave laser at the wavelength of 532 nm and the excitation power of 3 mW, measured in front of the microscope objective. In Fig. 3a, we plot the corresponding PL spectrum extracted from our device at various values of the in-plane moment recorded at 10K. The PL spectrum is widely dominated by the strongly parabolic cavity dispersion with an effective mass of $(4.2 \pm 0.1 * 10^{-5})$ $m_e$ ($m_e$ is the free electron effective mass), yielding highly photonic exciton polaritons with a photonic fraction (Hopfield coefficient $|C|^2$) of 0.967 at $k_\parallel=0$. Due of the high photonic fraction at $k_\parallel=0$ we can use the linewidth of the lower polariton branch to estimate the experimental Q-factor which amounts to 650. As the sample temperature is increased to 150 K (see Fig. 3b), the exciton resonance successively redshifts towards the cavity mode, and the system approaches the full spectral resonance at $\mathbf{k}_\parallel=0$, giving rise to well-pronounced upper and lower polariton modes. In order to extract the coupling strength of the GaAs-Wannier-Mott excitons, we fit our experimental data by the standard coupled oscillator model describing the normal mode coupling of exciton and photon

$$\begin{bmatrix} E_{\text{ex}} & V/2 \\ V/2 & E_{\text{ph}} \end{bmatrix} \begin{bmatrix} \alpha \\ \beta \end{bmatrix} = E \begin{bmatrix} \alpha \\ \beta \end{bmatrix} \qquad (1)$$

where $E_{\text{ph}}$ and $E_{\text{ex}}$ are photon and exciton energies, respectively, and $V$ the exciton-photon coupling strength. The eigenvectors yield the Hopfield coefficients for the exciton and photon fractions of the polariton states. The result of this modelling is shown in Fig. 3a and Fig. 3b along with the experimental data. Here, the red solid lines depict the polariton resonances, the dashed black line represents the empty cavity mode which was slightly shifted compared to the simulation in Fig. 2b) based on nominal structural parameters to account for processing and growth inaccuracies, the dashed green line is the exciton energy as a function of the in plane wave vector $k_\parallel$. Both fits independently yield a coupling strength of $V=9.0$ meV, which is in good agreement with the data from literature discussing similar Tamm-structures with embedded GaAs QWs [25].

We will now characterize the coupled $MoSe_2$-GaAs-cavity device, by studying the energy-momentum dispersion relation at the spot position corresponding to the TMDC monolayer location. This time the device is excited via a pulsed laser (82 MHz) at the wavelength of 745 nm and the excitation power of 500 µW. Fig. 4a) shows the PL spectrum taken without substantial spatial filtering, which we record on the monolayer. In this configuration both regions, with and without monolayer, contribute to the recorded signal. The spectrum clearly features two distinct branches with a similar (yet not identical) curvature. While the high energy branch has been assigned to the lower polariton mode, which is collected from the periphery of the monolayer, the lower energy branch reflects the influence of the monolayer exciton in the system. This can be directly verified by applying tight spatial filtering, which yields a PL spectrum solely dominated by the low energy branch (see Fig 4b). The tight spatial filtering is achieved by placing a pin hole in the real-space image of the PL signal. In order to approach the zero detuning regime in our coupled hybrid device, we again increase the temperature of the sample, as shown in Fig. 4c (spectra taken without strong spatial filtering) and Fig. 4d (with strong spatial filtering). The redshift of the GaAs excitons, as well as the $MoSe_2$ resonance leads to a decrease of the curvature of the polariton resonances, and we can clearly capture the characteristic polaritonic dispersion relation. In order to describe the eigenenergies of the newly emerging resonances, we extend our coupled oscillator model to the case of three oscillators:

$$\begin{bmatrix} E_{ex_1} & 0 & \frac{V_1}{2} \\ 0 & E_{ex_2} & \frac{V_2}{2} \\ \frac{V_1}{2} & \frac{V_2}{2} & E_{Ph} \end{bmatrix} \begin{bmatrix} \alpha \\ \beta \\ \gamma \end{bmatrix} = E \begin{bmatrix} \alpha \\ \beta \\ \gamma \end{bmatrix} \qquad (2)$$

Again, the three Hopfield coefficients quantify the admixture of QW- and monolayer-exciton ($|\alpha|^2; |\beta|^2$) and cavity photon $|\gamma|^2$. In order to fit the experimental data, we fix the coupling strength of the GaAs QW excitons with the cavity mode as 9.0 meV (see above), and describe the temperature dependent cavity modes at $\mathbf{k}_{||} = 0$ with $m_{cav} = (4.2 \pm 1 * 10^{-5})\ m_e$ (see above). The temperature dependent excitonic energies are described by the Varshni formula [26,27] and set as constant parameters for the fit function, which reveals a coupling strength of 20.0 meV for the MoSe$_2$ valley exciton. This coupling strength is in a good agreement with the literature [21,26].

While clear PL spectra can only be observed for the lowest hybrid polariton branch, we can identify the frequencies of the remaining two hybrid branches in the reflectivity spectra. Even though the combination of very tight spatial filtering and the highly asymmetric Tamm-structure yields only weakly pronounced resonances in reflectivity, we still can observe signals of three dominant resonances which coincide with the three hybrid polariton branches, in agreement with the coupled oscillator model (Fig. 4e).

These observations allow us to explain the emerging optical resonances in our device as three hybrid polariton modes in the collective strong coupling regime between the embedded GaAs QWs and the MoSe$_2$ monolayer. At the temperature of 140 K and normal incidence, this branch admixes a fraction of 8.95 % GaAs exciton, 15.5 % TMDC monolayer exciton, and 75.6 % cavity photon at $\mathbf{k}_{||}$=0 (Fig. 4f). Further information about the admixture of the exciton species for various temperatures are described in Supplementary Note 2 and Supplementary Fig 2.

We will now address the polariton distribution along the dispersion branches, Fig. 4d. We shall describe the energy-dependent polariton distribution function within the thermodynamic approach [22]. This simple model assumes perfect thermalization of the exciton-polariton gas. The PL from the polariton states $E_i(k)$ can be found as

$$I(k,E) \sim \sum_i \frac{|C_{\text{ph}}^i|^2 \exp\left(-\frac{E_i(k)}{k_B T}\right)}{\left(E - E_i(k)\right)^2 + \Gamma_{\text{ph}}^2} \quad (3)$$

Here, we assume the Boltzmann distribution of our quasiparticles: $N_i \sim \exp(-E_i/k_B T)$, where $N_i$ and $E_i$ denote i-state population and energy, respectively, and $k_B$ is the Boltzmann constant. We assume that the emission stems from the photonic mode (related to the photon Hopfield coefficient $|C_i|^2 = |\gamma_i|^2$) only and it is broadened in energy according to the Lorentz distribution.

$\Gamma_{\text{ph}}$ is the broadening of the photonic mode and the *i*-index spans over three polariton branches. The calculated dispersion relation is plotted in Fig. 5 for the temperature of 140 K. The qualitative agreement between theory and experiment is excellent, and, equally important, our model also explains the absence of the PL signal from the middle and upper polariton branch, which is due to the insignificant photonic fraction in the middle polariton branch and a weak thermal occupation in the upper branch.

**Discussion**

In conclusion, we have evidenced the formation of hybrid exciton-polaritons in the collective regime of strong coupling between Wannier type of excitons in GaAs quantum wells, strongly bound valley excitons in a MoSe$_2$ monolayer and cavity photons in a Tamm-plasmon-polariton device. We observe the three characteristic hybrid polariton resonances, and explain their occupation by a thermodynamic model. Our work manifests the first successful observation of this new kind of quasi-particles, and paves the way towards a manifold of applications of hybrid exciton-polaritons. While the monolayer excitons in principle are extremely robust and allow to maintain the spinor degree of freedom in a thus far unprecedented manner, the GaAs Wannier QW excitons are highly sensitive to external magnetic and electric fields, and strongly support lasing and condensation phenomena in the strong coupling regime.

We also believe that the long-living exciton reservoir provided by the GaAs excitons will strongly alter the relaxation and scattering dynamics compared to polaritons which are solely based on monolayer excitons. While direct current injection in monolayers has been demonstrated [28], including a vertical p-i-n junction in our device would allow us to operate our hybrid polariton device similarly to a standard III-V light emitting diode or vertical emitting microcavity laser.

Methods

**Sample design and fabrication**

The sample was designed by transfer matrix calculations, where the plasmon-polariton resonance was tuned to match the A-exciton and trion resonance of the MoSe$_2$ monolayer and the exciton resonance of the GaAs QWs. The bottom structure consists of an epitaxially grown DBR with 30 pairs of AlAs/Al$_{0.25}$Ga$_{0.75}$As layers (62.5nm/55 nm thickness, respectively, corresponding to a central stopband wavelength of 750 nm) and a 112 nm thick AlAs – layer on top with an embedded stack of four 5 nm thick GaAs QWs with 10 nm thick AlAs barriers in between. The bottom structure is capped with a 63 nm thick GaInP layer. The stopband ranges from 710 nm to 790 nm depending on the in-plane wavevector, and the GaAs – QWs emit at 1.658 eV. The MoSe$_2$ monolayer was mechanically exfoliated onto a polymer gel film (polydimethylsiloxane) and was then transferred onto the structure. 80 nm of PMMA were deposited by spin coating onto the structure. Finally, a 60 nm thick gold layer was thermally evaporated onto the sample.

**Experimental Setup**

We used an optical setup in which both spatially (near-field) and momentum-space (far-field) resolved spectroscopy and imaging are accessible. The sample temperature could be varied between 4.2 K and room temperature by using a Helium flow cryostat. PL is collected through a 0.65 NA microscope objective for the pure QW – polariton measurements and through a 0.42 NA microscope objective for the hybrid polariton measurements to enable strong spatial filtering, and directed into an imaging spectrometer with up to 1200 groves/mm grating via a set of relay lenses, projecting the proper projection plane onto the monochromator's entrance slit. The system's angular resolution is ~ 0.03 µm$^{-1}$ (~ 0.2°) and its spectral resolution is up to ~ 0.050 meV with a Peltier–cooled Si-CCD as detector.

Correspondence and requests for materials should be addressed to Christian Schneider (christian.schneider@physik.uni-wuerzburg.de).

**Figure**

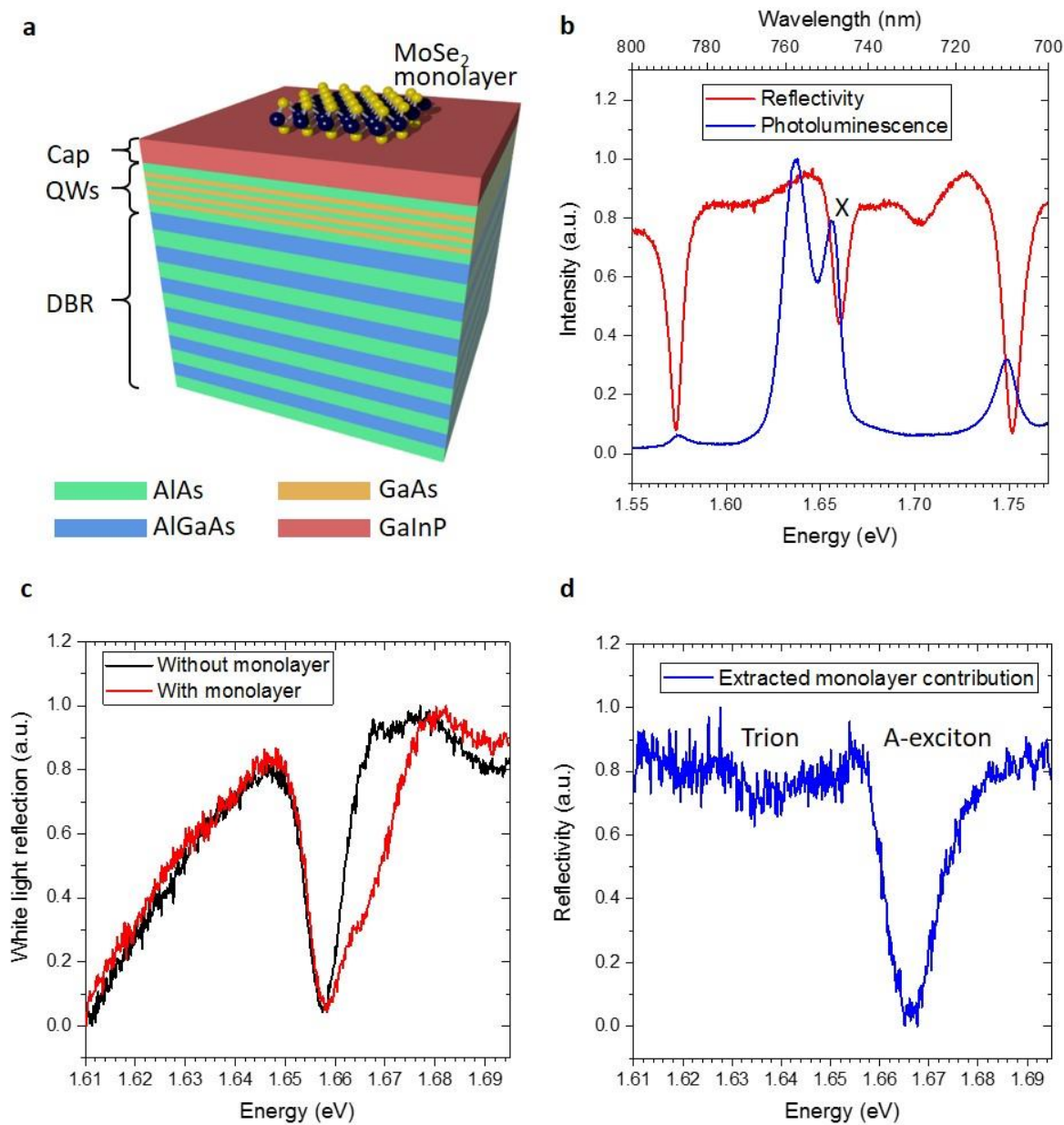

**Figure 1 | Base structure. a**, Schematic illustration of the epitaxially grown base structure with a mechanical exfoliated MoSe$_2$ monolayer on top of the GaInP cap. The Bragg wavelength of the bottom AlAs/AlGaAs DBR and the GaAs/AlAs QWs are designed to be resonant to the MoSe$_2$ A–exciton. **b,** Photoluminescence and reflectivity spectra of the structure without flake. The peak at 749 nm named X corresponds to the excitonic electron – heavy hole

transition at the gamma – point of the GaAs QWs (see Supplementary Note 1, Supplementary Fig. 1), which matches the absorption resonance of the stop band. The reflectivity spectrum is dominated by a stop band which ranges from 710 nm to 790 nm with a calculated reflectivity of over 99.9 % from 740 nm to 765 nm. **c**, White light reflection of the structure with and without the MoSe$_2$ monolayer. **d,** Absorption of the MoSe$_2$ – monolayer by norming the on–flake reflection to the off–flake reflection shows a strong absorption at the A-exciton energy, 1.666eV, and a weak absorption at the trion energy, 1.634 eV.

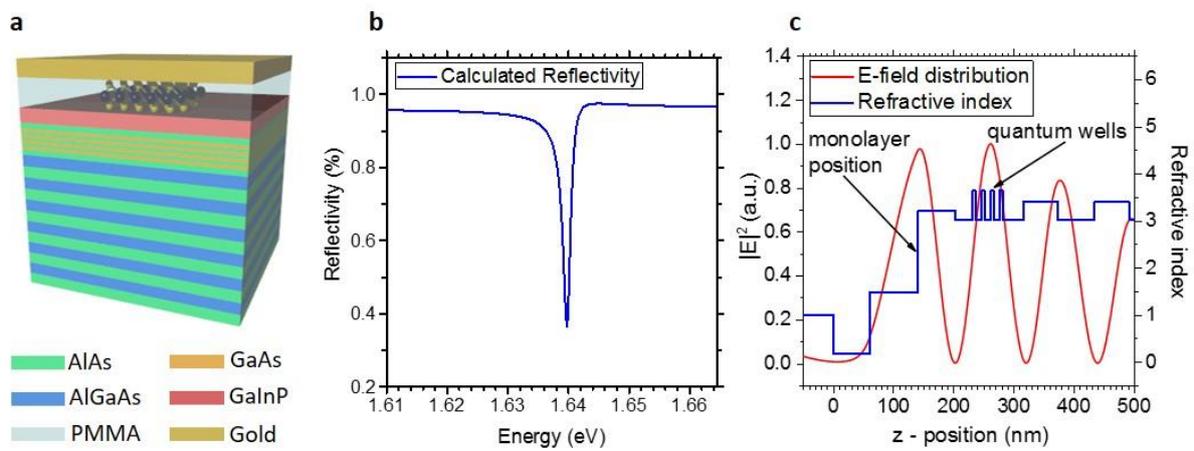

**Figure 2 | Tamm – quantum well – monolayer hybrid device a**, Schematic illustration of the Tamm-plasmon device with the embedded GaAs QWs and the MoSe$_2$ monolayer. **b,** Reflectivity spectrum calculated by the transfer matrix method, which yields the theoretical Q–factor of 1095. **c**, Layer sequence of the top part of the Tamm structure represented by the corresponding refractive indices (blue profile) and the simulated field distribution of the resonant mode (red) within the Tamm structure showing maxima at the QW and monolayer positions.

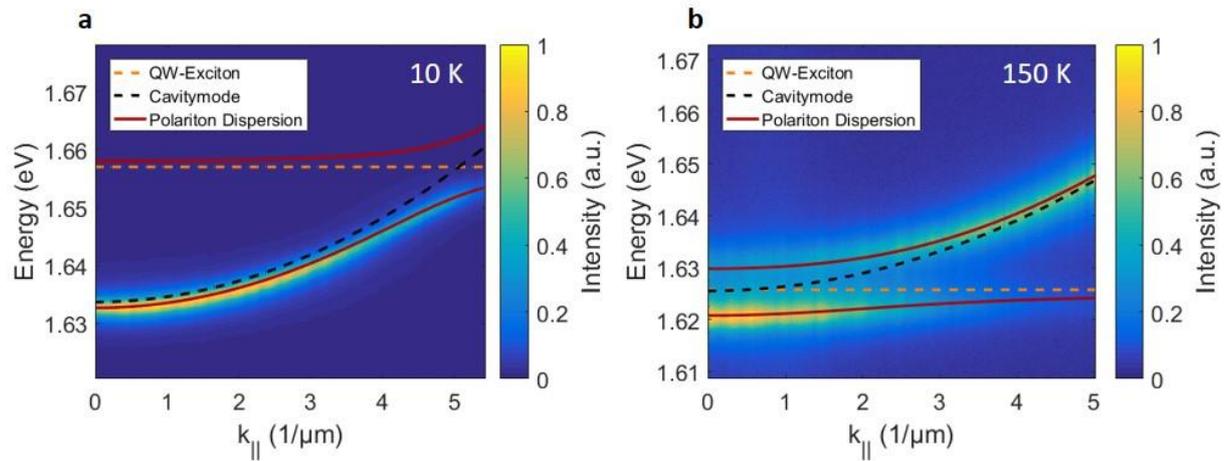

**Figure 3 | GaAs quantum well - polaritons. a,** Angle resolved photoluminescence measurements of the Tamm device without the monolayer at 10K. The dashed yellow line represents the QW exciton energy, the dashed black line corresponds to the cavity mode and the red line the shows the calculated polariton dispersions for the upper and the lower polaritons. **b,** Corresponding measurement at 150K, yielding the exciton-polaritons at zero detuning.

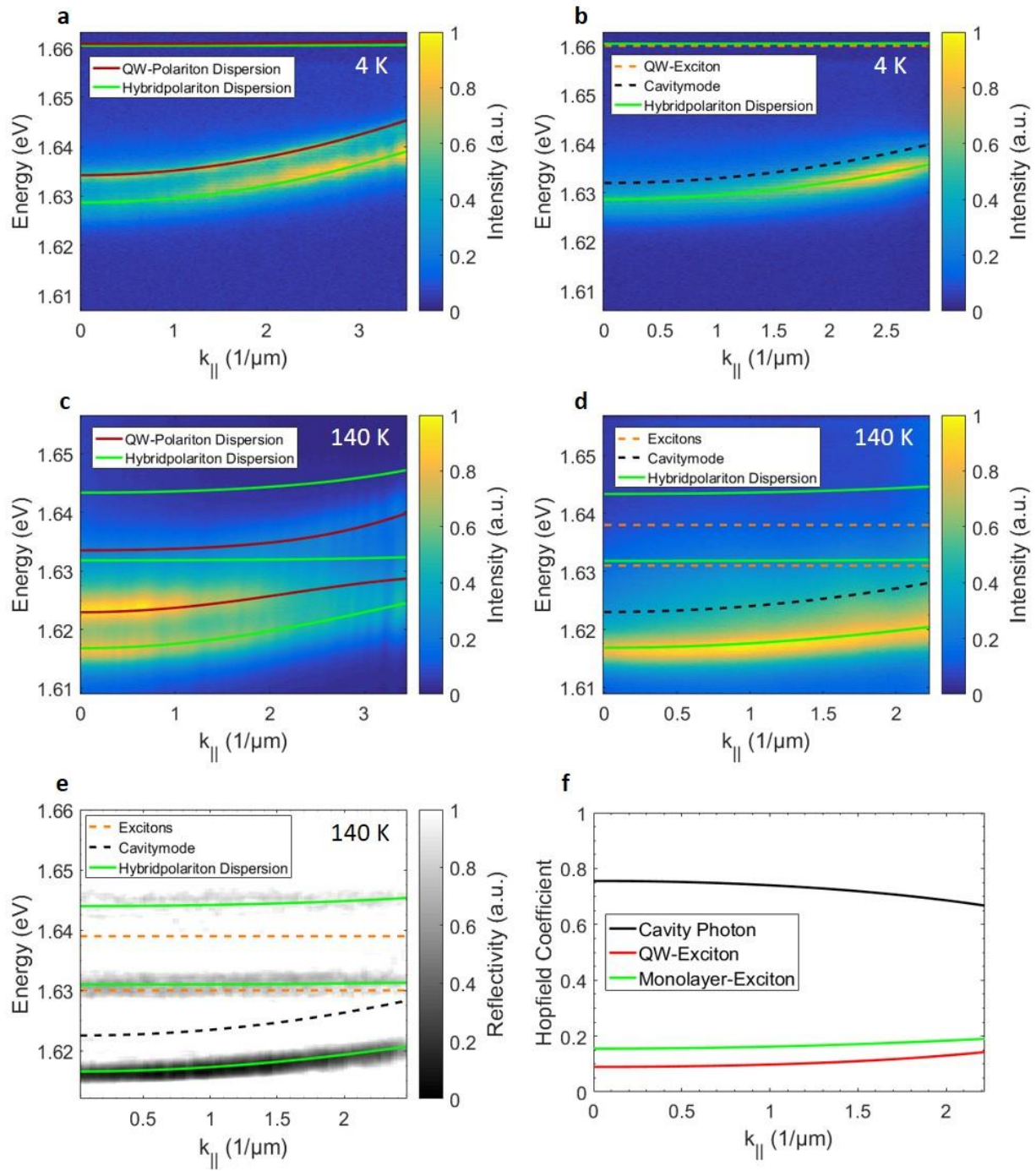

**Figure 4 | Hybrid polariton dispersion relation in a hybrid Tamm microcavity containing a monolayer of MoSe$_2$. a,** Angle resolved photoluminescence measurement at the Tamm device with the monolayer at 4K. The red line represents the calculated polariton dispersion and the green line shows the hybrid polariton dispersion for a slightly wider cavity (ca. 1 nm corresponding to 3 meV energy shift) at the flake position. **b,** Same dispersion measurement as a) but with strong spatial filtering at the flake position. The energy of the QW exciton is

represented by the dashed yellow line and the cavity mode at the flake position by the dashed black line. The MoSe₂ exciton energy of 1.666 eV is not shown on this chart. **c,** PL dispersion measured at 140K with the same color coding as before. **d,** Same measurement as c) but with strong spatial filtering at the flake position. The energy of the MoSe₂ exciton is shown by the orange dashed line. **e,** Angle resolved reflectivity measurement with strong spatial filtering at the flake position. **f,** Calculated Hopfield coefficients for the lower hybrid polariton branch.

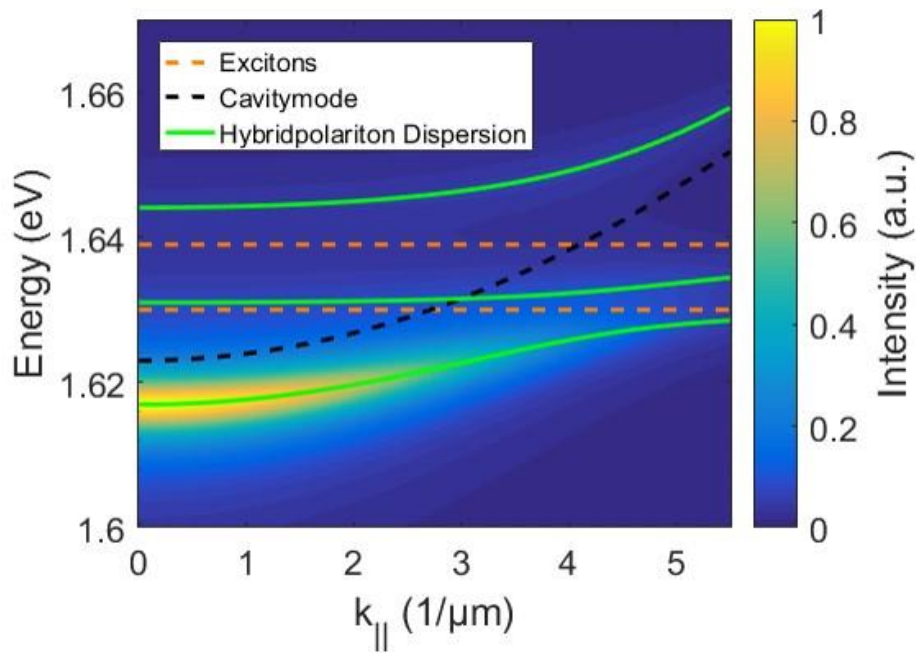

**Figure 5 | Simulated hybrid polariton dispersions of the Tamm device.** Occupation numbers of the hybrid polariton states at 140K are obtained using a thermodynamical approach. The parameters used for the simulation are obtained from fitting the data shown in Fig. 4d with the coupled oscillator model. The color coding is the same as used before.

Acknowledgement: This work has been supported by the State of Bavaria. C.S. acknowledges financial support by the European Research Council (unLiMIt-2D project). AK acknowledged the support from the HORIZON 2020 RISE project CoExAn (Grant No. 644076). S.H and A.K acknowledge funding by the EPSRC.